# Mirage : a new iterative Map-Making code for CMB experiments

D. Yvon[1,2] and F. Mayet[1,3]

[1] CEA-CE Saclay, DAPNIA, Service de Physique des Particules, Bat 141, F-91191 Gif sur Yvette Cedex, France  
[2] Fédération de Recherche APC, Université Paris 7, Paris, France  
[3] LPSC Grenoble, CNRS/IN2P3 and Université J. Fourier, 53 avenue des Martyrs, F-38026 Grenoble cedex, France

January 12, 2005

**Abstract.** A major goal of CMB experiments is to obtain highly sensitive CMB maps in order to extract Spherical Harmonic Power Spectrum (SHPS) and cosmological parameters with unprecedented accuracy. We present a new map-making code (Mirage), based on a two-phase iterative algorithm, involving low frequency drift treatment, Butterworth high-pass filtering and conjugate gradient method. This work was strongly motivated by Archeops CMB experiment data analysis. We believe that Archeops was a good test bench for the future Planck Surveyor data analysis, and Mirage was designed in order to be used for Planck data processing with minimal work. A strong feature of Mirage is that it handles experimental problems in data, such as holes in data stream, bright sources, and galaxy side effects, without jeopardising speed. The other advantage is its processing speed, allowing to run Monte Carlo simulations of Archeops data processing on a single processor workstation overnight. Algorithms are explained. Systematic effects on SHPS are investigated on various simulated data, including typical Archeops observational systematics. The code is now available at http://www-dapnia.cea.fr/Telechargement/tel_mirage.php

## 1. Introduction

Current and forthcoming CMB anisotropy experiments, both space borne (Planck[1], MAP[2]) and balloon borne, e.g. Archeops[3] [Benoît *et al.* 2003a, Benoît *et al.* 2003b, Benoît *et al.* 2003c], should allow for a major breakthrough towards the understanding of the early Universe, in particular through the estimation of cosmological parameters with unprecedented accuracy.

Amongst various steps of the CMB data analysis pipeline, from raw data treatment to angular power spectrum and eventually cosmological parameters extraction, the map-making phase enters as a crucial one, which should be handled with great care, in order not to induce systematics in the map, such as low frequency striping or galaxy side effects.

Most map-making methods rely on the assumption that the instrumental noise is gaussian, to assure that the "optimal" map is a maximum likelihood estimate[4] and stationary, or at least piecewise stationary.

The basic formalism of map-making algorithm has already been described in great details (e.g. [Tegmark 1997, Hamilton 2003]). Basic relations are recalled hereafter for the reader's convenience. Let $\mathbf{t}$ be the true sky convolved by the instrumental lobe, assumed to be symmetric, $\mathbf{n}$ the instrumental noise, $\mathbf{A}$ the pointing matrix, and eventually $\mathbf{d}$ the observed data stream. Sky patterns are stored in $N_{pix}$ pixels. Assuming that the Time-Ordered Data (hereafter TOD) depends linearly on the map allows to write the fundamental relation as :

$$\mathbf{d} = \mathbf{At} + \mathbf{n} \quad (1)$$

As $\mathbf{d}$ and $\mathbf{n}$ involve $N_{sp}$ samples, the pointing matrix $\mathbf{A}$ is a $N_{sp} \times N_{pix}$ matrix. The map-making method, if linear can be expressed (eq. 1) as : $\mathbf{s} = \mathbf{Wd}$, where $\mathbf{s}$ is the reconstructed sky map and $\mathbf{W}$ is the matrix describing the linear method. When computing the *error map*, one gets : $\mathbf{s} - \mathbf{t} = [\mathbf{WA} - \mathbf{I}]\mathbf{t} + \mathbf{Wn}$ with $\mathbf{I}$ being the identity matrix. It can be underlined that the property $\mathbf{WA} = \mathbf{I}$ is highly desirable, because, if true, the method is unbiased and the error map does not depends on the sky signal.

This allows, given $\mathbf{W}$ and a reliable noise model, to statistically *subtract* known noise contributions in the analysis.

The simplest map-making method is the *coaddition* :

$$\mathbf{W}_{coad} = [\mathbf{A}^t \mathbf{A}]^{-1} \mathbf{A}^t$$

The operator $\mathbf{A}^t$ projects the timeline onto the sky map pixels, the operator $\mathbf{A}^t \mathbf{A}$ counts the sample directed toward each pixel. This method is very simple, fast, and satisfies the property $\mathbf{W}_{coad} \mathbf{A} = \mathbf{I}$ and has been shown to minimise the error map variance for white noise. Unfortunately, in most ex-

---

*Send offprint requests to*: yvon@hep.saclay.cea.fr  
*Correspondence to*: yvon@hep.saclay.cea.fr

[1] http://astro.estec.esa.nl/Planck  
[2] http://map.gsfc.nasa.gov/  
[3] http://www.archeops.org  
[4] It can be noticed that even if the noise is not gaussian the optimal map has the lowest residual variance.

periments, the noise is not observed to be white. Defining $\mathbf{N} = <\mathbf{nn}^t>$ as the noise covariance matrix, it has been shown in [Tegmark 1997] that the optimal method is given by

$$\mathbf{W} = \left(\mathbf{A}^t \mathbf{N}^{-1} \mathbf{A}\right)^{-1} \mathbf{A}^t \mathbf{N}^{-1} \qquad (2)$$

This method satisfies the relation $\mathbf{WA} = \mathbf{I}$, minimises the chi-squared noise towards the reconstructed map, and is the maximum-likelihood estimate of true sky $\mathbf{s}$ if the probability distribution of $\mathbf{n}$ is gaussian. It has been shown in [Tegmark 1997] to be *lossless*, i.e. to retain all CMB information initially contained in the timeline. This algorithm (MADCAP [Borrill 1999]) has been successfully applied to several CMB data sets, such as COBE [Bennett *et al.* 1996], MAXIMA [Hanany *et al.* 2000] or BOOMERranG [de Bernardis *et al.* 2000]. Nevertheless, inverting the full matrix scales like $N_{pix}^3$, where $N_{pix}$ is the number of pixels in the map, thus limiting the interest of this method to rather *small* number of pixels in maps.

In the context of current and forthcoming experiments, for which the data set are extremely large ($N_{pix} = 10^7$, $N_{sp} = 8 \times 10^8$ for Archeops and $N_{pix} = 5 \times 10^7$, $N_{sp} = 10^9$ for Planck), several map-making codes have been developed and published, involving either a direct ($\mathbf{A}^t \mathbf{N}^{-1} \mathbf{A}$) inversion (MADCAP [Borrill 1999]), or an iterative inversion algorithm [Prunet 2000, Wright 1996, Natoli *et al.* 2001, Doré *et al.* 2001, Dupac *et al.* 2002]. FFT which scales only as $N_{sp} \log N_{sp}$ was the fundamental tool of these fast iterative map-making algorithms [Wright 1996] and is used intensively in Mirage. Recent implementations of some of the quoted algorithms improve even more this scaling to $N_{sp} * \log L$ where L is the useful bandwidth of the noise, a constant factor.

The Mirage map-making presented in section 3 is based on a two-phase iterative algorithm. To begin with, a quick estimation of the sky map is evaluated in the first phase, in which the low frequency drift is removed. This map is then used as an input for the second phase, a conjugate gradient method.

## 2. Observational constraints

When applying a map-making method to real data, several items must be carefully taken into account.
First, data streams do have holes that should be filled properly. They may be induced by low level data treatment, such as removal of glitches, i.e. high energy cosmic rays (mainly protons) hitting the bolometer.
Instrumental noise in cryogenic experiments often presents low frequency systematics whose knee frequency may exceed the scanning frequency. As shown in the following simulations, scan-synchronous low frequency systematics project on the map and cannot be minimised by the map-making algorithm. With ring-based scanning strategies (Archeops, Planck) annoying stripes show up on the sky maps.

These low frequency disturbances may be subtracted, if they turn out to be correlated to some other instrumental or observational data. If no useful decorrelation data is available two options are then available (unless a better idea shows up) : either to filter the data, suppressing the noise components along with the signal at the corresponding scales in the sky map, or to make a priori assumptions on the noise properties, and subtract the best estimate of this noise component. In the case of the MIRAGE algorithm, the filtering option has been chosen.
Then, the effect of the Galaxy must be taken into account. The Galaxy is, with the planets, the brightest object of the sky. When filtered, the rather large galactic signal, as compared to the expected CMB one, induces ringing on the timelines that distort the power spectrum at various angular scales.
As a matter of fact, Monte Carlo simulation is compulsory to understand data analysis systematic and correct for its effect. For instance, the MASTER method [Hivon *et al.* 2002] allows to calibrate the unwanted effects on the CMB power spectrum, of either the instrumental noise, high-pass filtering or any map-making induced alterations. This requires repetitive Monte Carlo TOD simulations, followed by the map-making of these data streams. Any useful map-making algorithm must thus be fast enough to allow for massive Monte Carlo simulations.
Next section presents the algorithm of the Mirage map-making code.

## 3. Mirage algorithms

### 3.1. General design options

We decided to develop two separated Map making algorithms, based on the same basic libraries and some specialised tools.
The first algorithm, named MirageDC, computes a sky map, without filtering the timeline. If the data does not display low frequency noise with knee frequency above the scanning (spinning) frequency, this is the algorithm to use.
The second, named MirageAC, computes a sky map, filtering the timeline using a parametrised Butterworth [Horowitz and Hill] high-pass filter. As expected, the low frequency features in the timeline and in the map are filtered out. Both algorithms work in two steps. The first step is a fast estimation of the sky map in order to minimise the work left to the iterative optimisation, which is the time consuming step of the process. The quick estimation map is thus saved and feeds the second step, the iterative map optimisation, based on a Conjugate Gradient iterator. The noise properties can be reliably extracted from the time ordered data, with no a priori. When the noise spectrum is complex, with a strong low frequency component, this two step method ensures faster and smoother convergence of the algorithms. The optimised sky map is saved, then subtracted from the input timeline: the best estimate of the noise spectrum is computed from this noise timeline and saved.
Finally speed was a constant concern in writing this code. The code is not, by far fully optimised, but all algorithms used scale with the samples number $N_{sp}$ as $N_{sp} \log N_{sp}$ or less, typical of Fast Fourier Transforms.These codes have been optimised for large sky coverage scanning strategies, such as Archeops and Planck, and obviously account for large signals induced by the Galaxy.

## 3.2. Gap filling

In order to compute a sky map, several variables are needed for each sample : the time (or sample number), the bolometer pointing Galactic latitude and longitude, and the bolometer signal, in physical unit on the sky. Most of the time the experiment provides also a quality indicator for the sample named flag. The flag is decoded according to the experiment : if the sample should not be kept, the pointing and bolometer signal are overridden to undefined value. Such undefined samples belong mostly to relatively short sets of data, but induce sharp discontinuities in the timelines that prevent to use Fast Fourier Transforms. To overcome this problem it is necessary to implement a gap filling algorithm that has to be fast and to preserve the spectrum of the timeline. Gap filling using linear prediction available in Numerical Recipes [Press *et al.*] provide satisfactory result, but scales as $N_{gap}^2$. Instead a simple mirroring technique has been chosen. Undefined bolometer signal samples are replaced by the mirror of the timeline taken at the last valid sample before the gap or the mirror of the timeline taken the first valid ample after the gap. Sharp discontinuity (step) between the last (first) filled sample and the following (preceding) timeline may then happen. Continuity is restored by adding to the filled samples a linear interpolation of the step amplitude. We then retain the mirror option of minimum variance after subtraction of the interpolation between the last valid sample before gap and the first valid ample after gap. The algorithm computation time scales linearly with the gap sample number. Additionally, using simulated noise timeline with spectrum of figure 2 with up to 20% undefined samples, the reconstructed noise spectrum after gap-filling was checked to be statistically compatible with initial noise spectrum, thus behaving satisfactory.

## 3.3. Conjugate Gradient Iterator.

The Map making equation written as $(A^t N^{-1} A)\tau = A^t N^{-1} d$, where $\tau$ is the vector of scanned sky pixels values to be computed, is a set of linear equations, with each observed pixel of the sky being a unknown. The Conjugate Gradient Method (CG) [Barrett *et al.*] is an iterative numerical method widely used for solving systems of linear equations. Following [Natoli *et al.* 2001], this method has been chosen. Computation speed is improved by two methods :
1) Given an estimated sky $s_0$ the map-making equation can be written as

$$(A^t N^{-1} A)(\tau - s_0) = A^t N^{-1}(d - A s_0)$$

This variable change, suggested by [Doré *et al.* 2001], when used with a reasonable initial sky guess, remove most of the signal from the remaining timeline $D = d - A s_0$ ($A s_0$ being the timeline computed when the estimated sky $s_0$ is *observed* through the pointing matrix). It allows to cut the range of numbers involved in the computation thus improving numerical stability of the algorithm. It also minimises sharp transitions on the timeline data, when the line of sight crosses bright objects such as the Galaxy. $D$ is now assumed to be dominated by noise and is used to compute the noise spectral density $\Xi$. Assuming the noise is stationary, [Tegmark 1997] explains how to a very good approximation :

$$N^{-1} D = F^t \left[ \Xi^{-1} F(D) \right]$$

where $F$ ($F^t$) is the Fast Fourier Transform (antitransform). Computation time scales as $N_{sp} \log N_{sp}$ and the large Matrix $N^{-1}$ does not need to be stored.

2) Convergence speed is improved by using a preconditioning matrix. A first choice of the preconditioning matrix is : $P = [diag(A^t N^{-1} A)]^{-1}$, where the diag operator reduces the $A^t N^{-1} A$ to its diagonal part, which can be approximated by : $P = \nu_{noise}(A^t A)^{-1}$, square matrix whose diagonal element value $P_{ii}$ is the noise variance divided by the number of measurements of the sky for the pixel $i$. The map making equation is solved in its final form :

$$P(A^t N^{-1} A)(\tau - A s_0) = P A^t N^{-1}(d - s_0)$$

The CG method requires to provide the initial residual $A^t N^{-1}(d - A s_0)$ and the operators working on the sky, $P$ and $A^t N^{-1} A$. Then a stopping criterion is required. If quick estimation algorithms are well designed, little work is left to the Conjugate Gradient optimization and we expect quick convergence. A first choice should be to stop the optimization when the optimization algorithm converges, limited by machine precision and often associated with small oscillations in the map residuals. But systematic errors on the Spherical Harmonic Power Spectrum (SHPS) due to sky pixelisation (Healpix pixel window functions, see sec. 5.1) turns out to be enhanced by the CG iterator and poor scanning strategy redundancy. This effect is difficult to correct for, if the iteration number changes from one timeline to an other. This is why we chose to stop the iteration at a predefined iteration number $N_{iter}$. In the following simulations we used $N_{iter} = 5$, as a working choice that minimizes computation time.

Given the pointing matrix $A$, operating $A$ and $A^t$ scales as $N_{sp}$, and the CPU time used by this Conjugate Gradient iterator scales as $N_{iter} N_{sp} \log N_{sp}$, thus being very efficient.

## 3.4. Classical "optimal" map-making : MirageDC

### 3.4.1. A fast timeline slow-drift substractor : SlopeKillerMask

The goal of the quick estimation algorithms is to provide quickly a rough but reliable estimation of the sky to feed the input of the Conjugate Gradient method. Quick maps have to be good enough to ensure that when subtracted from the timelines, the residual is dominated by noise.
SlopeKillerMask assumes that the observing strategy describes large circles on the sky. In particular, this is the scanning strategy chosen by Archeops and Planck. The most dramatic noise effect on timeline is low frequency noise (see Figure 3b), and the Galaxy. In the following low frequency noise is assumed to have negligible power above the scanning (spinning) frequency. Then SlopeKillerMask masks Galactic signal, computes the low frequency component of the timeline, subtracts it and then calculates the map by simple coaddition.
SlopeKillerMask first scans the timeline pointing, looking for the maxima in Galactic latitude and store them in memory. Data between these maxima are stored in set named *Circles*. For each circle, samples pointing toward a bright region of the galaxy are flagged and discarded: bright Galactic regions are defined as brighter than a user chosen threshold (typically 5 MJ/st) at a wavelength of $100\,\mu$m using SFD Galactic model [Finkbeiner *et al.* 1999]. We fit over the valid Bolometer signal sample values a straight line with time. The *Offset* is finally stored as the value of the fitted line, in the middle of the circle. This is a very robust estimation of the low frequency evolution of the timeline, provided the data has been properly flagged for glitches and cosmic rays, before the map-making process. If not, this algorithm can be made insensitive to glitches, by running the straight fit twice: after fitting the straightline, we compute the variance of the timeline minus the fitted line. We flag samples that are 5 standard deviations from the fit. Doing so, we found that moderate and large amplitude glitches as well as bright point sources were flagged with adequate efficiency. We then computed once more the straight line fit, and store the fitted value in the middle of the circle.

The list of Offset is then used to compute the estimated slow drift component of the timeline by simple interpolation in the list, and this slow drift timeline is subtracted from the bolometer signal. A simple upgrade would subtract a spline function of the offset list instead of interpolations. The resulting timeline $\mathbf{d}_{sub}$ is then projected on the sky $\mathbf{W}_{coad}\,\mathbf{d}_{sub} = \mathbf{s}_0$, giving the fast rough estimation. Computation time scales as the timeline sample number and is found to be negligible as compared to the rest of the map-making process.

### 3.4.2. Iterative Optimisation

The iterative optimisation of the map is a straightforward use of the Conjugate Gradient iterator described in sec. 3.3. It uses the input timeline and the rough sky estimation to compute the optimised sky map. 5 Conjugate Gradient cycles are used, and the residual modulus typically gets divided by 100. This is indeed the time consuming step. It scales as $N_{iter}N_{sp} \log N_{sp}$.

### 3.5. MirageAC

When low frequency noises happen to pollute data above the scanning frequency, the SlopeKillerMask algorithm does not produce adequate maps. MirageAC was designed to high-pass filter the timeline, to suppress noise as well as signal under the frequency cutoff. The fundamental building block of such a method is a filtering tool. We chose to filter in the Fourier space, taking care of edge effects, and used the gap filling software to prepare the timelines. The main drawback of AC filtering is related to the galaxy. The Galaxy center is very luminous compared to the CMB. As expected, high-pass filtering induced ringing around large features. The way out is to subtract overly luminous object from the timeline before filtering. But large low frequency noise prevents to make a good estimation of the bright signal before filtering. We designed an iterative algorithm to achieve a compromise.

### 3.5.1. Quick map estimation: Rough Filtered Map-Making

The input timeline is kept unmodified all along the process. A copy of the input timeline is high-pass filtered, and projected in a filtered sky map. The variance of the distribution of the filtered sky map is computed outside the Galaxy. Filtered sky map pixels are then saved above a threshold of 2.5 standard deviations in a bright object Map : $\mathbf{s}_B$.

Making an other copy of the input timeline, we first subtract the timeline obtained by observing the bright object Map: $\mathbf{A}^t\,\mathbf{s}_B$ . We filter the resulting timeline, project it, threshold it and add the result to the bright object Map.

We loop on the process four times. Then in the last loop, we do not threshold the filtered sky map, but we add it to the last bright object Map to compute the Rough Filtered Map, with noise.

This algorithm was shown to effectively remove low frequency noise and suppress ringing effect on the timeline and computed maps by a factor 20. Of course, the bright sky features (Galaxy) are high-pass filtered too, in a way that depends of the scanning strategy. If the scanning strategy is complex (Archeops), the bias induced on the maps can only be studied through simulations. This algorithm scales as $N_{sp} \log N_{sp}$, (dominated by the FFT transforms) and is thus significantly slower than the SlopeKillerMask algorithm and follows the same scaling as the iterative optimisation. An interesting alternative to this code is the *filtering without ringing* algorithm presented in [Amblard & Hamilton 2003]. Tests need to be performed to compare respective performances.

### 3.5.2. Iterative Optimisation in MirageAC

The iterative optimisation follows the same ideas. The Rough Filtered Map is *subtracted* from a copy of the input timeline. The resulting timeline, assumed to be dominated by noise is processed by the Conjugate Gradient Iterator, that compute a Correction Map. The optimised sky Map is the sum of the Rough Filtered Map and the Correction Map.

## 4. Simulated timeline data

In order to test and optimize the map-making method, simulation tools have been developed to generate realistic timelines.

### 4.1. Input Sky Maps

A CMB temperature anisotropy map is generated using CMBfast [Zaldarriaga *et al.* 2000] and Synfast [Gorski *et al.* 1998] codes. For this tests, *standard* model of Cosmology has been used, namely :

$\Omega_{\rm tot} = 1, \Omega_{\Lambda} = 0.7, \Omega_{\rm M} = 0.026, \Omega_{\rm B} = 0.04, h_{100} = 0.65$

A galactic map is generated using a galactic dust emissivity model. The extrapolation model from [Finkbeiner *et al.* 1999]

is used, in particular their best fit model. This map is extrapolated from the $100\,\mu$m emission and $100-240\,\mu$m flux ratio maps from IRAS and COBE-DIRBE data. To take into account a finite instrumental resolution, the input map is convolved with Gaussian beam, assumed to be 10' FWHM. Optical frequency dependance of sky signal (dust, CMB, ...) is integrated on optical filter assumed to be of top hat shape, with $\Delta f/f = 30\%$. Convolved CMB maps and galactic maps are then obtained and *observed*, given the pointing of the instrument, to obtain a *signal only* timeline.

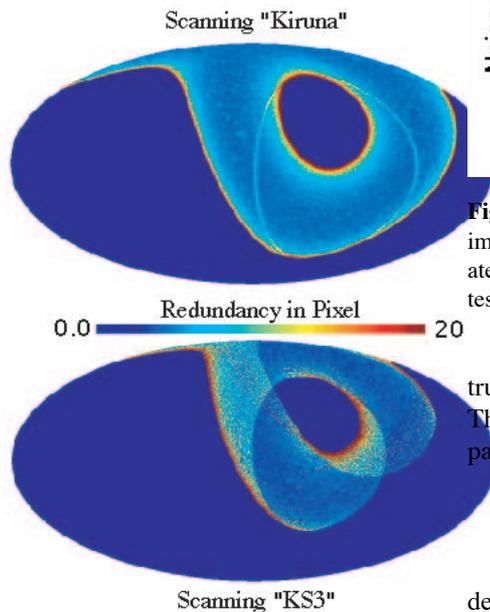

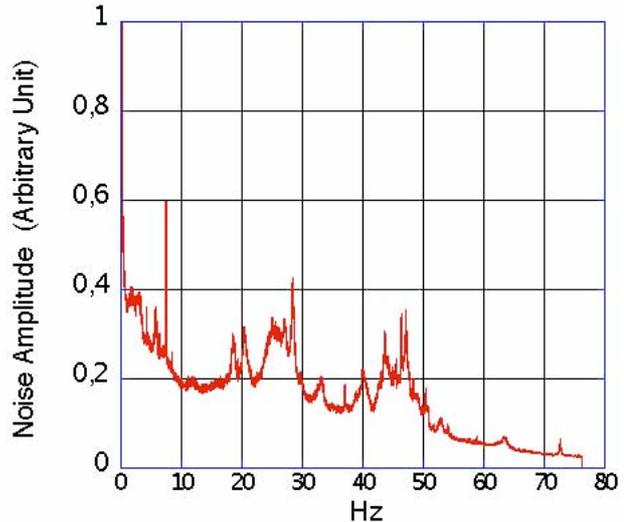

**Fig. 2.** Noise spectrum of what would be a strongly polluted experiment, displaying large and wide noise spikes, additional to moderate $1/f$ noise excess. This noise spectrum is used in simulation as tough test of the robustness of the algorithms presented in this paper.

**Fig. 1.** Mollweide projections of the two scanning strategies used in this study. Kiruna scanning strategy for the Archeops mission. **a)** The label Kiruna correspond to the theoretical sky coverage of a 25 hours Archeops Flight. This sky coverage is homogeneous with timelines cutting at large angle and no data hole. **b)** Sky coverage of the third Archeops Kiruna Scientific flight (KS3): a *real life* example.

### 4.2. Scanning strategies

Two scanning strategies are being used. Their Mollweide projections of sky coverage are shown on figure 1. The first one is a *theoretical* Kiruna 24 hour scanning strategy for the Archeops mission, limited to $9 \times 10^6$ samples, to allow quick tests on our workstation. This is a best case sky coverage : homogeneous, with timelines cutting at large angle and no data hole. The second is a copy of an Archeops third Kiruna Scientific Flight (KS3) bolometer pointing file, including data holes, and inhomogeneous sky coverage, that is an example of *real-life* data.

### 4.3. Noise Models

Our models assume that noise properties are stationary. Noise timelines are then added to the *signal only timeline* to make the input timelines of the map making. Two kinds of noise spectrum are being used :

The first one is the classical $1/f$ and white noise spectrum parametrised as following :

$$\text{NPd}(f) = e_n \times (1 + f_{\text{knee}}/f) \times f/(0.002 f_{\text{knee}} + f)$$

where $e_n$ is the high frequency *flat noise* power spectrum density and $f_{\text{knee}}$ is the $1/f$ knee frequency of the spectrum. The term $f/(0.002 f_{\text{knee}} + f)$ has been added to avoid the divergence of the integral of the Noise Power Spectral density at zero frequency. In the following, this noise spectrum is used unless specified.

The second model shown on figure 2 uses the noise spectrum of what would be a strongly polluted experiment, displaying large and wide noise spikes, additional to a moderate $1/f$ noise. Though such an experiment is not likely to detect much of the CMB signal, the heavy structure of the noise spectrum is a strong test of the robustness of the algorithms presented in this paper.

The Mirage code may then be tested with various physical inputs (CMB, galaxy), level of noise as well as various scanning strategies. The entire simulation tools are developed in C++ within the framework of SOPHYA [Ansari *et al.*].

### 4.4. Mirage algorithms working on simulated data

As a first test of systematics induced by our algorithms we compute noise free timelines by reading a Galaxy + CMB anisotropy sky map. We then process these timelines by Mirage algorithms and subtract the input sky map to produce difference maps. The results are shown at Figure 3. We see that SlopeKillerMask algorithm induces little distortion while MirageDC optimiser minimise somehow the galaxy brightest pixels. MirageAC quick estimation and optimiser do induce some residual ringing that distorts the galactic map, but this ringing is small enough to allow CMB anisotropy $C_\ell$ estimation, if we mask the galaxy area.

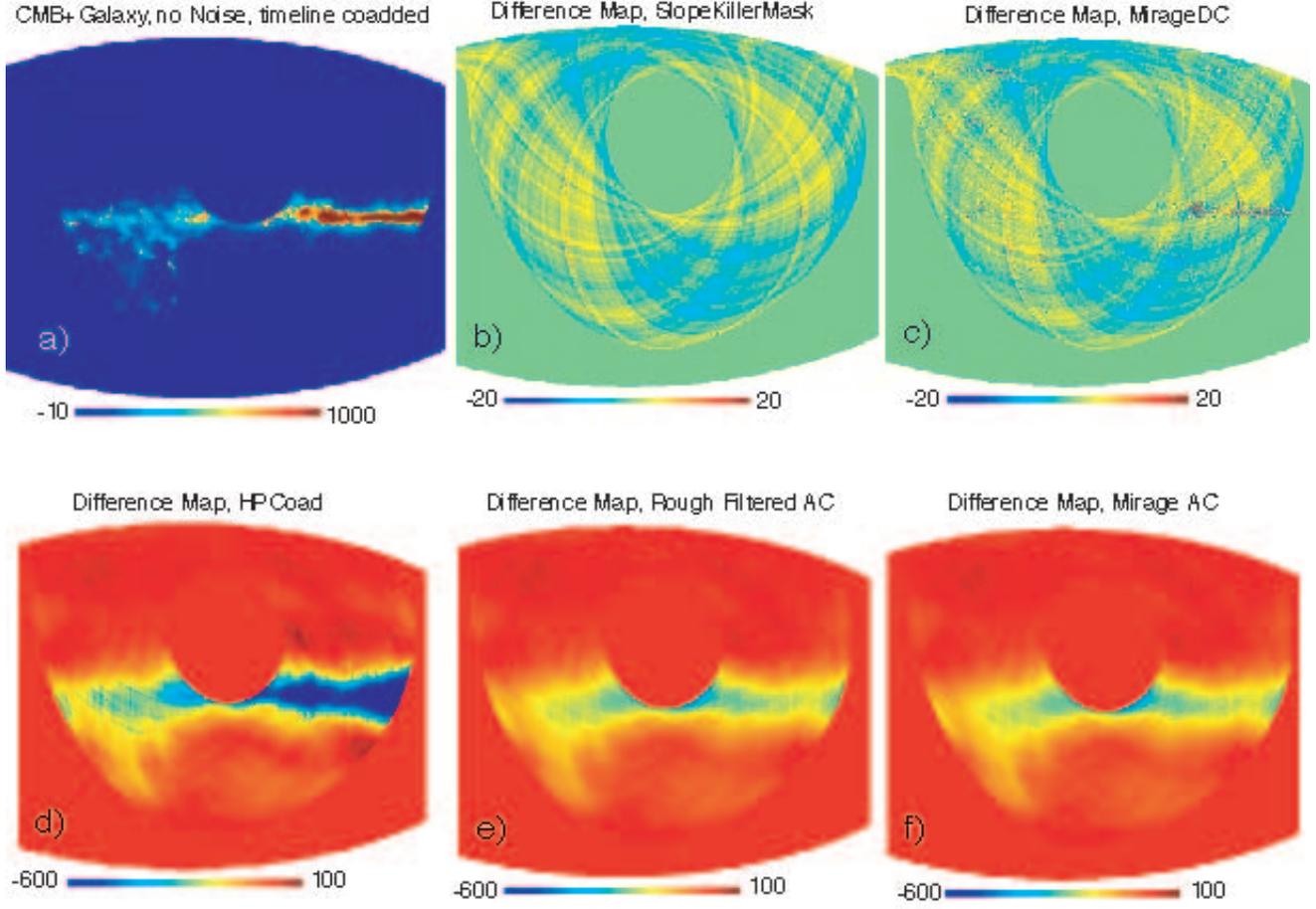

**Fig. 3.** Mollweide projections of sky maps computed by Mirage algorithms from simulated timelines : **a)** is the Signal (CMB anisotropy + Galaxy) map to be reconstructed ; **b)** to **f)** are difference maps: maps processed by Mirage algorithms running on simulated timelines with CMB + Galaxy signal only, Signal map subtracted. These maps show and sometimes enhance the systematics induced by the algorithms processing the timelines, with signals of large amplitude (Galaxy map brightest pixel's value is 13000). We observe that SlopeKillerMask **b)** induces very little distortion, while MirageDC **c)** tends to minimise the Galaxy brightest pixels. **d)** shows the difference map computed by high-pass filtering and coaddition of the timeline. Large systematic shows in and around the galaxy area due to the high-pass filter and its side effect ringing. **e)** and **f)** show the efficiency of MirageAC and the rough filtered map making algorithms to minimise distortion and ringing in maps. Still large distortions remain and MirageAC is of little use for mapping the Galaxy or other foreground, but turns out to be very useful for CMB $C_\ell$ estimation when data involves noise with high frequencies and strong 1/f features.

As a second test, Figure 4 shows Mollweide projections of sky maps computed by Mirage algorithms using noisy simulated timelines. Timelines have been generated using a Galaxy + CMB anisotropy map, $1/f$ and white noise. Mirage algorithms process these timelines and we subtract the input sky map to produce the difference maps shown. Figure 4a shows the difference map obtained if we use the simplest of all map-making algorithm: simple projection and averaging of measurements on sky pixels (coaddition) used on a timeline with noise knee frequency lower than the scanning redundancy frequency. The map is dramatically dominated by noise-induced lines following the scanning path named *stripes*. Figure 4b shows the difference map computed using the SlopeKillerMask algorithm to remove low frequency drifts followed by coaddition. SlopeKillerMask as well as MirageDC optimiser pro-duce maps with no more stripping, and the noise level is lower where the scanning frequency shows higher redundancy. (Figure 4c et 4d are difference maps computed from a timeline with noise knee frequency above the scanning redundancy frequency. Figure 4c shows the optimised map computed by MirageDC. It shows strong pollution by striping, due to the low frequency noise synchronous with observations that project on the sky. Figure 4d shows difference map processed by the optimised MirageAC output map. Please note the change in scales, and the noise amplitude variation with the scanning strategy redundancy.To remove striping seen in Figure 4c, we chose a MirageAC high-pass filter cutting frequency of one tenth of the noise knee frequency. The striping is dramatically reduced, though still noticeable. Rising the high-pass frequency allows to suppress stripes completely but increases at the same time filtering of the underlying CMB signal. Thus the choice of the filter cutting frequency is a trade off.

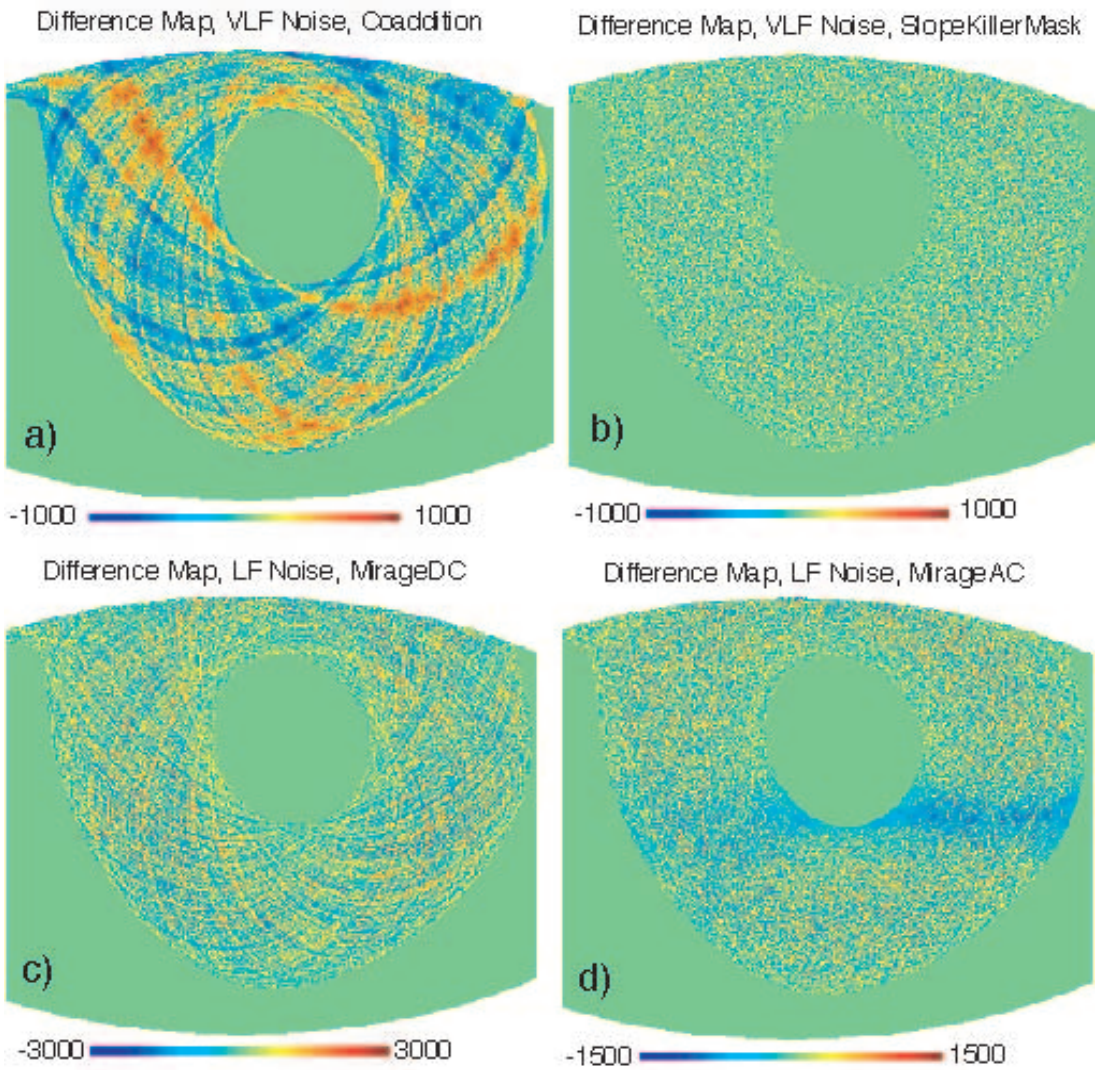

**Fig. 4.** Mollweide projections of Difference sky maps computed by Mirage algorithms from simulated timelines : The signal map simulates the CMB anisotropy and the Galaxy. **a)** to **d)** are Difference maps processed by algorithms running on simulated timelines with Signal and $1/f$ + white noise. VLF stands for Very Low Frequency: the noise knee frequency is below the scanning redundancy frequency. LF stands for Low Frequency: the noise knee frequency is chosen above scanning redundancy frequency. **a)** shows the results of simple coaddition on Very Low Frequency Noise. SlopeKillerMask and MirageDC **b)** algorithms are very efficient over very low frequency noise. On low frequency noise, MirageDC **c)** performs better than SlopeKillerMask, but fails as expected to remove stripping from maps. MirageAC algorithms **d)** are necessary to remove stripes if the timeline is contaminated by low frequency noise, but induce residual ringing around the galaxy area that will have to be masked for CMB $C_\ell$ estimation.

## 5. Efficiency and systematic errors on SHPS induced by Mirage algorithms

All data presented in the following are simulations. This work follows [Hivon *et al.* 2002], and then [Amblard & Hamilton 2003] within the Archeops collaboration where biases on SHPS induced by coaddition-based algorithms are studied in details. In this paragraph we explain methods used within Mirage optimisation algorithms to minimise its systematical errors, we quantify the algorithms efficiencies and residual biases. The SHPS is calculated simply by computing the $Y_{lm}$ transform of the analysed map using the AnaFast algorithm [Gorski *et al.* 1998] and then the Pseudo $C_\ell$. No attempt was made to de-correlate the $C_\ell$ values due to the limited sky coverage.

### 5.1. Sky pixelisation systematic errors

In the following, the Healpix pixelisation [Gorski *et al.* 1998] have been widely used, as well as associated algorithms. The first systematic error to enter when a map is used as an intermediate step before computing $C_\ell$ power spectrum is the well known "pixelisation" effect. Healpix pixelisation provides its analytically computed pixel window function that allows to correct the effect of finite map pixel size, in the evaluation of SHPS. It is found that an healpix resolution factor $N_{side} = 512$, or larger, need to be used to compute SHPS with acceptable distortion up to $\ell = 600$. It turns out that the Conjugate Gradient optimiser manipulate data in the form of pixelised maps at each iteration. We observed that the pixel window effect is amplified when the number of iteration increases. The natural remedy would be to pixelised the sky at a higher resolution. But in real life data (Archeops), as well as for our simulations, increasing the healpix resolution factor at 1024 would result in a typical pixel measurement redundancy of one. The large sky coverage, associated to the short flight duration prevents us to use healpix resolution factor larger than 512. Therefore, in order to be able to correct for the effect of pixel window function, we decided to run the CG optimiser for a user-defined number of iteration. In the following all maps were pixellised using a resolution factor $N_{side} = 512$. We then compare the additional distortions induced by the algorithms.

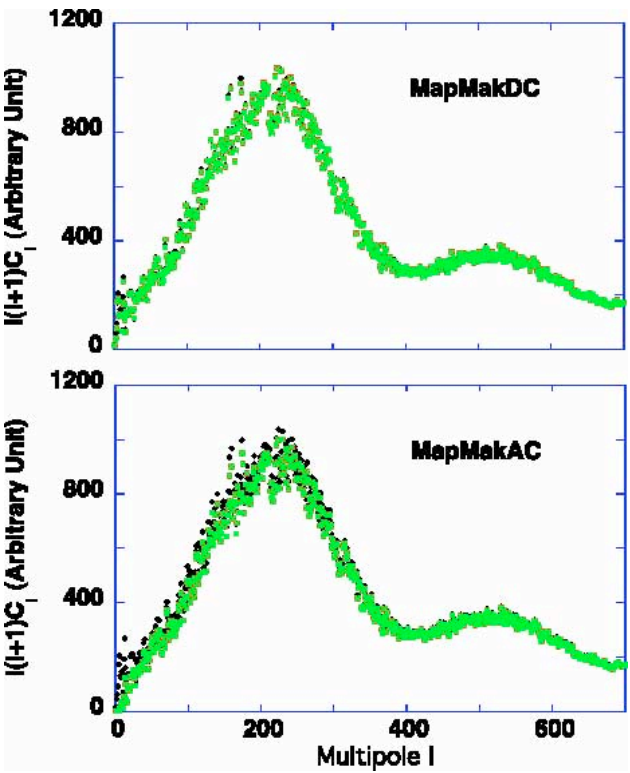

**Fig. 5. a) Upper panel.** Spherical Harmonic Power Spectrum of MirageDC computed maps. All the maps have been pixelised with Healpix resolution parameter $N_{side} = 512$. Black dots are the SHPS of the reference CMB map with pixels outside the observed sky suppressed. Red dots is SHPS of the map computed from a *CMB only* timeline using the SlopeKillerMask algorithm. Green dots is the SHPS extracted from same timeline with MirageDC iterative algorithm. No systematics is observed. **b) Lower panel.** Same for the MirageAC algorithms. Black dots are for SHPS of the input CMB map, red dots plot SHPS of the GalaxMapMak computed map, and green dots is the SHPS extracted from MirageAC optimised Maps.

## 5.2. MirageDC

### 5.2.1. Filtering function

In this paragraph, distorsions induced by the MirageDC are estimated, when used to process noiseless timelines. The input timeline is prepared as following: The CMB map is observed using a Kiruna 24 Hours scanning. No noise is added. We run the MirageDC and get 3 maps, a simple coaddition, the SlopeKillerMask quick estimation, and the iteratively optimised maps. These are, as expected, so similar that they can only be compared through their SHPS. Figure 5 shows the SHPS of the input convolved CMB map, over the SHPS of the 2 output maps. No distortion is seen on the SHPS.

### 5.2.2. Noise suppression efficiency

The three above mentionned map making algorithms have all been tested and found to be linear with respect to their inputs. The simplest way to quantify the noise suppression efficiency is then to run the Mirage on *Noise only timeline*. As a matter of fact, the lowest the noise spectral amplitude, the better the suppression (and hence the algorithm). In order to compare the efficiency of different algorithms, we ran Mirage on *Noise only timeline* using 3 kinds of noise spectra and 2 kinds of scanning strategies. The noise labelled VLF is a $1/f$ + white noise with knee frequency of 0.1 Hz. The noise spectrum labelled LF is a $1/f$ + white noise with knee frequency of 5 Hz. Finally the *bad* noise spectrum is the one presented in figure 2
The computed SHPS are shown on figure 6. Since the noise spectra are very different in amplitude, and the scanning strategies are of very different redundancy and sky coverage, the Y axis amplitudes should not be compared between different windows. The relevant information is the relative efficiency of the various algorithms.
Figure 6 shows the most relevant results of these simulations. We see at figure 6 a) and b) that the simple coaddition is the worst of the algorithm. If noise excess happen only at very low frequency, a simple *Slope Killer* followed by a coaddition is very efficient at suppressing noise, and conjugate gradient iterative optimisation does not improve the results. If noise excess happens to have a knee frequency above the spinning frequency, MirageDC Conjugate Gradient optimisation significantly improves the results, for both scanning strategies. Finally if the noise spectrum display strong peaks spreaded all over the spectrum the CG optimisation stays the best of MirageDC algorithms.
Now if we look closely at the reconstructed SHPS spectrum by MirageDC CG algorithm, we notice that the spread in 3 lines of the SHPS is present, though at a much lower level than with simpler algorithms.

### 5.2.3. Is MirageDC the final word ?

As shown above MirageDC is found to be efficient. The worst the noise spectrum is, the more efficient is the algorithm compared to simpler algorithms. But, when the low frequency noise knee frequency happens to exceed the scanning strategy spin-ning frequency, excess spin synchronous noise projects on the sky producing striping on the maps and the line pattern of the SHPS. Though MirageDC mathematically solves the map-making problem by minimising the variance of noise in the map optimally, it is not satisfactory for the physicist, because the line pattern of the pseudo-$C_\ell$ spectrum will be a big issue in our attempts to recover the primordial CMB SHPS at all multipoles. Some other input, prior or idea is required to improve our map-making process. Noticing that a strong low frequency noise will overwhelm the CMB SHPS at low frequency, we chose to high-pass filter the timelines, sacrificing the low multipole SHPS information, in order to remove the striping: this is MirageAC. The hope is to recover the high multipole SHPS information.

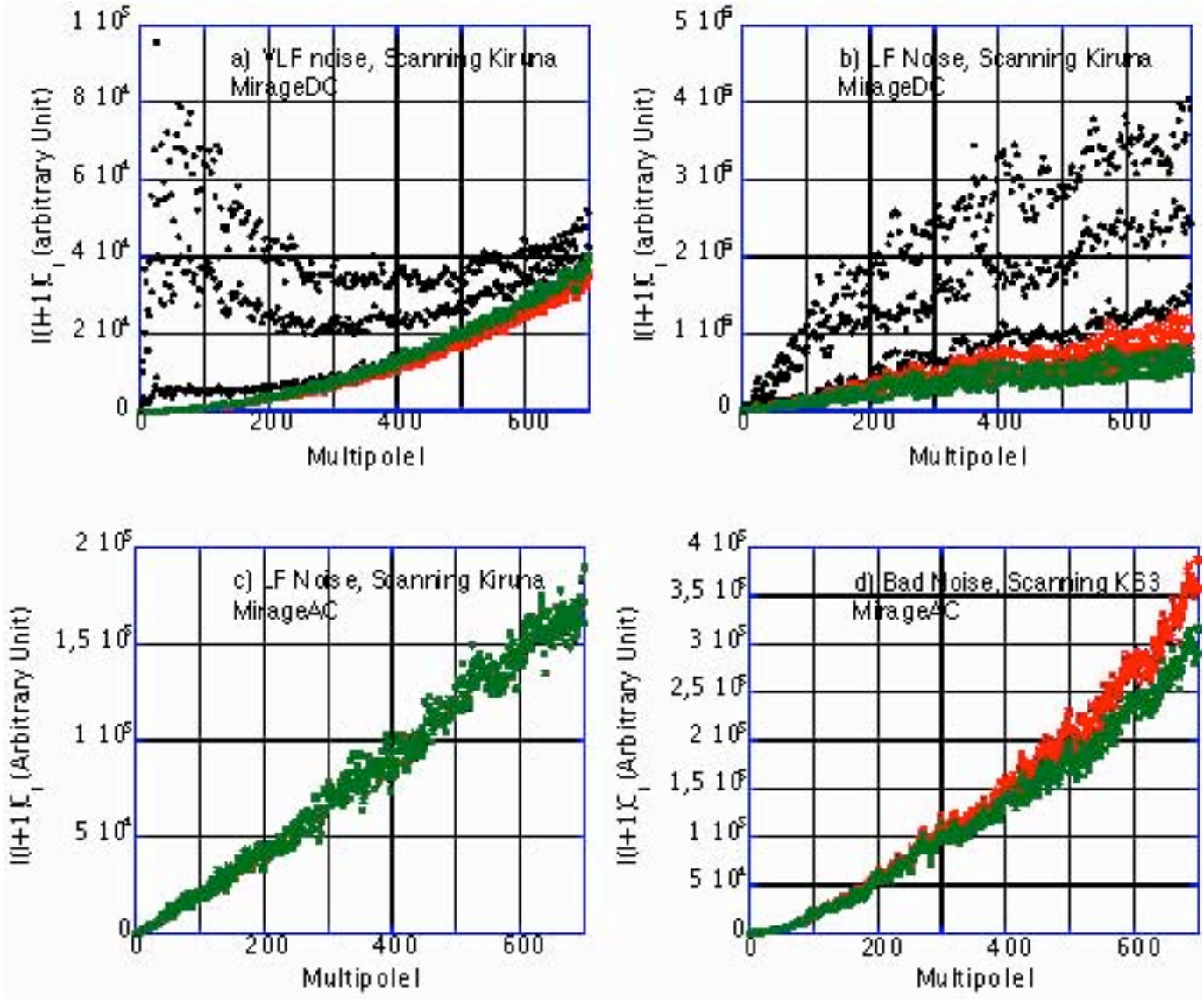

**Fig. 6.** SHPS computed from maps calculated with MirageDC and MirageAC algorithms. The timelines used included no signal, were generated from 3 differents noise spectra, and projected on the sky following two scannings strategies called Kiruna and KS3 (See fig. 1). **a)** The black lines are computed by simple coaddition, the red is the result of SlopeKillerMask algorithm, the green is the output of iterative optimisation using Conjugate Gradient method. We see that for noise peaked at below scanning redundancy, both MirageDC algorithms do the job perfectly. For knee frequencies low frequency noise above the scanning frequency **b)**, MirageDC optimiser improves the results compared to SlopeKillerMask algorithm as expected, but residual stripes in maps split the SHPS curve in 3 branchs. **c)** and **d)** SHPS computed from MirageAC maps algorithms working on the same low frequency noise timelines. The black lines are computed by high-pass filtering and coaddition of the timelines, the red is the result of GalaxMapMak algorithm, the green is the output of iterative optimisation using Conjugate Gradient method. **c)** Mirage AC improve a lot the SHPS compared to **b)** and we notice that the 3 algorithms show very similar performances for low frequency dominated noise. But if the noise spectrum and/or the scanning strategy are more complex **d)**, iterative optimisation of maps improves the noise rejection.

### 5.3. MirageAC

#### 5.3.1. Filtering Function

For this test we use the same timelines as those prepared for sec. 5.2.1. We run the MirageAC, with high-pass cutting frequency of 0.3 Hz, and a 4 order Butterworth filter. (Figure 5b) presents the SHPS obtained from the 3 output maps as well as the SHPS computed from the coaddition of the same timeline. It can be seen that the 3 algorithms induce very similar systematics on SHPS. Systematics induced by high-pass filtering the timelines is significant at low multipole values, and need to be evaluated and corrected from simulations.

#### 5.3.2. Noise Suppression efficiency

Map-making AC is used to process the same timelines as in sec. 5.3.1. Some results are presented on Figure 6c and 6d. We observed that when the 1/f + white noise knee frequency is lower or close to the High-Pass cutting frequency, Mirage AC efficiently suppress 1/f noise contribution to SHPS as expected. The 3 outputs maps provides the same SHPS spectrum even for LF noise, if the scanning strategy is homogeneous (Kiruna). But if the scanning strategy is inhomogeneous (e.g. Archeops), conjugate gradient optimisation does improves the result. Finally, if the noise spectrum is peaky, the Conjugate Gradient optimisation improves the result at high multipole values (6d). Archeops data analysis showed that MirageAC is very effective, when the noise spectrum rises at high frequencies: this happens when deconvolution of bolometers thermal constant low-pass filtering is necessary.

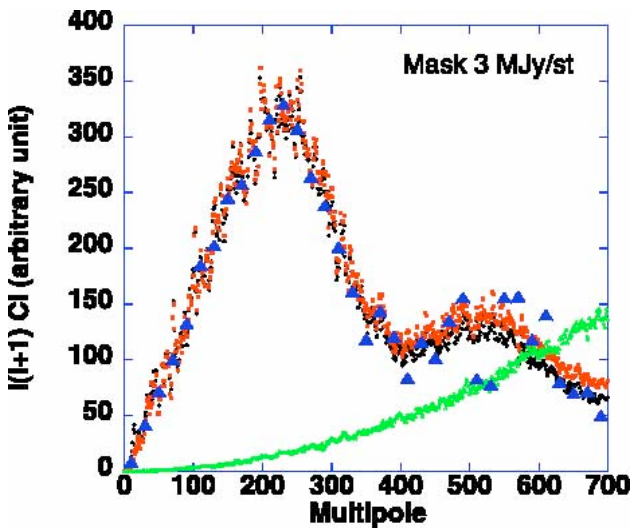

**Fig. 7.** SHPS of 4 maps from Mirage AC processed timelines. Simulated timelines have been generated according to text. Maps pixels pointing toward IRAS 100 m flux is larger than 3 MJ/sr have been put to zero. Black dots are the SHPS of *CMB only* map, pixelised at same resolution, and with the same sky coverage. Red dots presents SHPS of *Signal* map, Noise1 map subtracted. Green triangles shows the SHPS of the Noise1 map, divided by 10. The recovered SHPS follows closely the CMB SHPS with a small excess at multipoles larger that 400. Blue triangles presents rebinned *Signal* Map SHPS, SHPS of "Noise2" map subtracted.

#### 5.3.3. MirageAC and the Galaxy, validation of the method

The Galaxy is the largest and the brightest object of the Sky in the few 100 GHz bands. This reality induces two additionnal problems for evaluation of SHPS using maps of the sky.
The easiest to understand is that after a large signal, any high-pass filter rings over some period of time. The ringing feature biases the map over an area that has to be masked before SHPS estimation. This effect was taken care from the start in the MirageAC algorithm.

The second effet is a bias in the noise spectrum evaluation [Amblard & Hamilton 2003]. The power radiated by the galaxy varies continuously as the scan moves through it. For noise estimation, the quick sky signal estimation is subtracted from the noise timeline. This sky signal estimation is pixelised with finite resolution, the same as the output maps ($N_{side}$ = 512 for this work). When the scan crosses a quickly varying region of the galaxy, the timeline samples digitizes a fast changing signal, and the sky estimation averages this signal over the pixel size. When subtracted, the resulting timeline is contaminated by spurious bipolar features of pixel size, that consequently distorts the noise spectrum. This biased noise spectrum drives the Conjugate Gradient optimisation algorithm to minimise some of the CMB features. To correct this behavior, we state that the slope of Galactic signal is strongly correlated to the Galaxy intensity. Before computing the timeline noise spectrum, sky estimation subtracted, we erase the samples pointing toward a galactic area brighter that a threshold of 10 MJy/sr. The timeline is then restored with the gap-filling algorithm, and the *corrected* noise power spectrum is eventually computed.

A final test has eventually been performed to check for additional bias and validate Mirage AC. The simulated timelines were prepared, using for the sky signal the sum of a simulated CMB map and of the galaxy map observed with Archeops scanning pattern. Noise was generated using a $1/f$ + white noise spectrum. The same noise generation was used for the Sky + Noise1 (*Signal*) timeline and the noise only timeline *Noise1*. In addition, a second noise timeline *Noise2* was generated with the same noise statistical properties. High-pass filter cutting frequency and slope were set at 0.3 Hz and 4 poles. In the output maps, the areas where IRAS 100 $\mu$m flux was larger than 3 MJ/sr have been put to zero.

MirageAC processed the 3 optimised maps. In order to test for additional bias in map-making, we first subtract from Signal Map the Noise1 map, and compute the SPHS. Though unrealistic in real life, this is a tough test of the method bias, because we expect to recover input signal without noise nor distortion. Figure 7 shows the results: black dots are the "CMB only" SHPS, processed with MapMakAC, pixelised at same resolution, and with the same sky coverage. Red dots are the recovered SHPS. Green triangles show the SHPS of the Noise1 map, *divided by 10*. We see that the recovered SHPS follows closely the CMB SHPS with a small excess at multipoles larger that 400. No excess power is seen, when processing a noiseless Galaxy+CMB timeline. This excess power is found to depend on the noise amplitude and is due to a small error when the noise spectrum is extracted from the data [Amblard & Hamilton 2003]: the quick-estimation maps are contaminated by some residual noise. When subtracted from input timeline, the noise spectrum is then underestimated. This bias on noise amplitude is small (2% for this simulation), can be easily corrected by Monte Carlo simulations and is found to be negligible compared to the statistical noise as shown in figure 8 (blue triangles). In this way, we subtracted Noise2 SHPS from the Signal Map SHPS. The reconstructed SHPS (best estimation of experimental CMB SHPS) is noisy, and needs to be rebinned (blue triangles). We see that at multipoles larger than 400, where the bias become significant, the scattering of recovered SHPS around the CMB SHPS is much larger.

## 6. Conclusion

Two algorithms have been developed in order to optimise the sky maps computed from noisy timelines. Conjugate Gradient optimisation is efficient at minimising noise in timeline. CG is effective on timelines displaying complex noise spectrum and inhomogeneous sky coverage. If data is contaminated by low frequency systematics with a knee frequency larger then the scanning frequency, MirageDC, though being effective at reducing noise on the SHPS, does not succeed in removing stripes on the computed maps. This problem is shared by all "Optimal" map-making software [Borrill 1999, Prunet 2000, Wright 1996, Natoli *et al.* 2001]. The MirageAC algorithm does the job beautifully, at the cost of suppressing the power spectrum at low multipoles. This effect must be corrected using Monte Carlo simulations. In this way, MirageAC can be seen as an improvement of the MASTER map-making method [Hivon *et al.* 2002] (high-passed coadded timeline), with *optimal* processing of medium and high frequency noise features (lines and bumps in noise spectrum).

Both algorithms proved to handle experimental data efficiently, including data holes and bright objects, and to be of very low cost in CPU time. Mirage does not support nor needs parallel processing. On a single processor Compaq Alpha 667 MHz workstation, MirageDC processes a $10^7$ samples timeline in 20 s for the quick estimation, and 40 s per CG iteration.

Mirage (AC algorithm) has been used for a second estimation of Archeops CMB power spectrum using a cross-spectrum method [Tristram *et al.* 2004].

*Acknowledgements.* The authors wish to thank, the ARCHEOPS collaboration for using, criticising and motivating the development of Mirage software, C. Magneville and the SOPHIA team for discussions and providing support on the mathematicals librairies used by Mirage, J.-Ch.Hamilton, J.-P. Pansart, C. Yeche, J. Rich, R.Teyssier, J. Macias-Perez, M. Tristram, F-X. Désert and the referee for meticulously reading and commenting this manuscript, O. Doré and J.-Ch. Hamilton for introducing us to the subject in the early times of this work.